# Influence of fictive temperature and composition of silica glass on anomalous elastic behaviour

R Le Parc[1], C Levelut[1], J Pelous[1], V.Martinez[2] and B Champagnon[2]

[1] Laboratoire des Colloides, Verres et Nanomatériaux, CNRS/UMR5587, Université Montpellier II, cc 69, 34095 Montpellier cedex, France

[2] Laboratoire de Physico-chimie des Matériaux Luminescents, Université Lyon 1, CNRS/UMR 5620, 69622 Villeurbanne Cedex, France

leparc@lcvn.univ-montp2.fr

**Abstract**
In order to point out the influence of thermal history (fictive temperature) and OH content on the elastic properties of silica glass, we have performed high resolution in situ Brillouin spectra of $SiO_2$ glass from room temperature to the supercooled liquid at 1773K across the glass transition. The well known anomalous increase of elastic modulus in the glassy state and in the supercooled liquid regime is observed. No change of slope in the elastic moduli of silica appears as a characteristic of glass transition, on the contrary to what happens in various other glasses. We show that thermal history has a weak effect on elastic moduli in the glass transition regime for silica glass. The effect of water content in silica glass is more efficient than the fictive temperature effect and gives larger changes in the amplitude of elastic modulus for the same thermal dependence. A singular decrease above 1223K is also observed in the shear moduli for hydrated samples. Different models explaining elastic properties temperature dependence in relationship with frozen-in density fluctuations or with the structure are discussed.

PACS

## 1. Introduction

Silica glass is one of the most widely studied materials both theoretically and experimentally. On one hand silica glass is often considered as an archetype for amorphous material. On the other hand some properties of silica glass behave anomalously with temperature: density, thermal expansion, internal friction and elastic moduli show different behaviours as compared to crystals or other glasses [1]. Upon compression some anomalous variations are also reported for elastic properties: elastic moduli decrease from room pressure up to 2-3 GPa and increase for higher pressure [2].

*1.1. Thermal dependence of the elastic properties*

In this paper, our interest will be focused on thermal variation of elastic properties in different silica glasses. For most glasses, alkali silicates for example, elastic moduli decrease slightly in the glassy state and decrease with an important slope in the supercooled liquid state [3-5]. High temperature elastic moduli in silica have been measured by Brillouin light scattering [6-10] and ultrasonic measurements [11]. Both types of measurements have shown that elastic moduli increase with temperature, strongly in the glassy regime above 1000K and slightly in the supercooled liquid regime. Another anomalous temperature variation has also been reported at low temperature: elastic modulus is characterized by a minimum around 83K [12] related to a broad acoustic loss peak [13, 14] also seen at hypersonic frequencies [15-17].

Several explanations either based on phenomenological approaches or based on structural origins have been proposed for anomalous elastic behaviour in silica.

Among phenomenological approaches, Anderson reported a possible link between elastic properties and thermal expansion coefficient [18]. In silica glass, the thermal expansion coefficient presents two particularities: between absolute zero and room temperature, silica glass exhibits a negative thermal expansion coefficient α [1, 12] and at higher temperature, although positive, the thermal coefficient of silica glass is 10 to 30 times smaller than in alkali silicate glasses [12, 19]. Anderson postulates that the modulus thermal dependence can be considered as the sum of a pure volume term at constant temperature and a pure temperature term at constant volume. Thus the elastic moduli M can be expressed as a function of volume and temperature following equation 1.

$$dM/dT = \alpha V (dM/dV)_T + (dM/dT)_V \quad (1)$$

At first, the second term in equation is positive. Secondly, the first term is very small for silica glass because of the small thermal expansion. Finally, the sum of the two terms in silica glass is positive. In a usual glass as the thermal expansion is not small, the first term is not negligible and give rise to a negative $dM/dT$. This explanation based on the idea that a positive temperature coefficient of moduli would be expected every time α is small. However $GeO_2$ and $BeF_2$ glasses have both rather large thermal expansion coefficient although they show positive elastic moduli thermal dependences [12].

Several models giving interpretation of the anomalous elastic moduli for silica glass based on elastic inhomogeneities [20] or on the coexistence of two different structural states [21, 1] will be discussed in the following.

*1.2. Impurities and thermal history dependence of the elastic properties*

Measurements of elastic properties from the literature have been performed on silica glasses from different origins and proved that silica glasses from different sources or with different thermal histories are characterized by different measurable properties [19].

The thermal history of a glass can be characterized by the fictive temperature $T_F$, a quantity introduced by Tool [22]. The fictive temperature is defined as the temperature at which the liquid structure is frozen when cooling down through the glass transition. In the glassy state $T_F$ is constant and $dT_F/dT = 0$. In the glass transition range $dT_F/dT$ varies from 0 to 1 and in the equilibrium state $dT_F/dT = 1$. The fictive temperature characteristic of a given sample, can be changed by isothermal treatment at a temperature $T_a$ in the glass transition range. During this annealing at $T_a$, properties of the material are going to evolve with time, in order to reach the equilibrium corresponding to this temperature $T_a$. During the treatment, $dT_F/dT$ slowly relaxes from 0 to 1. After a treatment time long enough $\tau_a$ the properties do not evolve anymore, the sample is stabilized. Sample can then be quenched from $T_a$ to room temperature in order to freeze the new equilibrium and to obtain a sample with $T_F = T_a$ provided the quenching rate is fast enough. When the temperature of the treatment is in the lower part of the glass transition interval, the relaxation process is usually described by the term "aging".

Influence of thermal history on many properties in silica glass has been studied in various papers [19, 23-25, 26-32]. Some properties have been found to behave anomalously as compared to most other glasses. For example, when fictive temperature is raised, density and refractive index increase in silica glass [19, 23, 33] whereas they decrease in most usual silicate glasses [35].

Silica glasses are usually classified in four categories as regard to their synthesis process [19]. Class I contains silica glasses produced from natural quartz by electric fusion under vacuum or under an inert gas atmosphere. They contain a few OH ppm but a high concentration of metallic impurities (Al or Na). Class II contains silica glasses produced from crystalline quartz powder by flame fusion, thus OH content is about 150-400ppm but metallic impurities content is low. Types III and IV are composed by glasses produced by hydrolysis of $SiCl_4$, they contain about 100ppm of chlorine and about 1000ppm of OH for the first one and 0.4ppm OH for the second one. In this paper we will work on samples from class I, III and IV and a sol-gel glass prepared from a heat treated aerogel.

Several papers [36, 37, 38] have pointed out several modifications of the properties of silica glass induced by aluminium, fluorine chlorine and OH impurities. In this paper we will mainly concentrate on the influence of OH content measured for some properties in the literature [19]. The influence of OH content and fictive temperature are somehow intricate: increasing OH concentration also induce a decrease in viscosity [39], therefore the glass transition interval and/or fictive temperature are also modified by OH content.

The aim of this paper is to extend available data for silica glass using samples of various origins, thermal treatment and OH content. Acoustic properties will be deduced from in-situ temperature Brillouin shift measurements performed from room temperature up to high temperatures, extending above the glass transformation range. Elastic moduli will be deduced taking into account the variations of many different parameters used for the calculation of moduli (refractive index, density…). We also want to focus on the influence of thermal history and OH content on relaxation processes around Tg.

## 2. Experimental results

*2.1. Experimental devices*

Brillouin light scattering corresponds to coherent inelastic scattering associated to the interaction of light with long wavelength elastic waves (phonons) due to propagating thermal vibrations in condensed materials. It reflects the dynamics of the elastic excitations, which is determined through the viscoelastic properties of the scattering medium. Thus, Brillouin scattering gives access to the elastic properties at small wave vectors (compared to ultrasonic experiments) and at high frequencies (typically 30 GHz).

The Brillouin scattering measurements of transverse and longitudinal sound velocities at hypersonic frequencies have been performed using a high resolution Fabry Perot spectrometer [40, 41]. The incident light was the 514.5 nm line of a single mode argon ion laser. The incident power of the laser was about 600mW. A two- or four- pass plane Fabry-Perot, with a free spectral range equal to 75 or 56 GHz and finesse equal to 40 is used as a monochromator. The frequency corresponding to the maximum transmission of this filter is matched with the frequency of the Brillouin line. The resolving unit is a spherical Fabry Perot (SFP) interferometer with a free spectral range of 1.48 GHz and a finesse of 50. The total contrast is about $10^9$. The measurements of the longitudinal Brillouin shifts in samples have been performed in backscattering geometry in a Linkam TS1500 heating device designed to work from 335K to 1773K. Transverse shifts have been measured with right angle geometry in a furnace with optical windows built by Hermann Moritz (power maximum 4kW, maximum temperature 1773K).

The Brillouin peak position is measured between two consecutive orders for the spherical Fabry-Perot of the elastic light and thus the Brillouin shift determination is given including an error on the determination of the peak position but also an error on the determination of the free spectral range of the FSP between two consecutive orders of the elastic light. Hypersound velocity $V_{L/T}$ (L for longitudinal and T for transverse sound velocity) can then be calculated from the Brillouin shift $\nu_{L/T}$, from the refractive index n at a wavelength $\lambda_0$=514.5nm, and from the scattering angle ($\theta$=180° for backscattering) according to equation 2 :

$V_{L/T}= \Delta\nu_{L/T} * \lambda_0/(2n*\sin(\theta/2))$ (2)

Due to the aperture of the collecting lens, the broadening line is much more important with 90° angle geometry measurement (transverse shift measurements) than with backscattering configuration (longitudinal shift measurements).

Considering all those source of errors, the total uncertainty on the determination of sound velocity is about 5m/s for longitudinal sound velocity (in the backscattering geometry) and about 30m/s for transverse sound velocity respectively (with 90° angle geometry). An error on refractive index is discussed in the following (paragraph 2.2).

Longitudinal modulus M (or $C_{11}$) and Shear modulus G (or $C_{44}$) can be calculated from longitudinal sound velocity and transverse sound velocity and sample density (equation 4).

$M=\rho V_L^2$ (3)
$G= \rho V_T^2$ (4)

Brillouin scattering measurements correspond to very high frequencies, often considered as infinite frequencies compared to other determination of mechanical properties. Adiabatic compressibility (at infinite frequency) $\beta_\infty$ can be calculated from the inverse of the compression modulus K determined from Brillouin measurements (equation 5).

$K= M-4/3G$ and $1/K=1/(M-4/3G)=\beta_\infty$ (5)

*2.2. Parameters for the elastic properties determinations*

Determinations of samples density and refractive index are necessary in order to deduce mechanical properties from Brillouin data (equation 2, 3, 4). For this study refractive index and density variations with fictive temperature and OH content have to be considered.

For the silica glass samples studied, containing a few ppb OH, presented in the subsection 2.3., density has been measured using a pycnometer in toluene [42]. An increase of the density $\rho$ by about 0.2% with $T_F$ is observed in the [1373-1773K] interval. The general evolution can be approximated roughly to a linear law (equation 6).

$\rho = 10^{-5}T_F + 2.1869$ (6)

Data from literature show that silica glass density increases with increasing fictive temperature in the range 1373-1673K [19, 23, 43]. An increase of density (and thermal expansion) with quenching rate, has also been observed using molecular dynamic simulation [44].

Data from Brukner [19] show a maximum in the density for $T_F$ in the range 1673-1773K, but this result has been attributed to the impossibility of reaching high fictive temperatures with air or water quench [45]. In the case of the samples studied, these have been proved to be well stabilized by other characterizations (Raman and IR) [46, 47]. The density difference between the samples respectively heat treated at 1673 and 1773K (table 1) is within the experimental error, then it appears difficult to draw a conclusion about a potential maximum density for our samples. In a recent paper from Sen [48], a density minimum for a fictive temperature equal to 1223K is reported for silica glass. Such a minimum has been observed for our samples around 1373K [49], but the relaxation time at 1223K in dehydrated samples containing a few Al ppm is very long and the stabilization of the samples (during one month heat treatment) is to be verified.

The density temperature dependence can be estimated from the linear thermal expansion coefficient taken as equal to $0.5*10^{-6} K^{-1}$. The volume thermal expansion $\Delta V/V=\Delta\rho/\rho$ is then equal to $1.5\ 10^{-6} K^{-1}$ and for $\rho =2.203$, the density variation in the range of measurements, $\Delta\rho=3.304*10^{-6}$, can be neglected.

As for the evolution of the density with OH content, an increase by $0.96*10^{-6}$ for each 1wtppm of water regardless of $T_F$ has been reported [42]. In a paper from Fraser [50] an increase of density for a constant fictive temperature is estimated to $0.5 *10^{-6}$ per wt ppm.

Refractive index versus fictive temperature $n(T_F)$ has been measured in different papers [19, 23, 33]. It increases almost linearly with increasing fictive temperature from 1273 to 1673K and can be related to density variations, as shown in reference [23]. As for the density, above 1673K the existence of a refractive index maximum is also debated. Different slopes can be found for $n(T_F)$, according to wavelength and silica glass composition. For the analysis of our data, we used for our samples, an interpolation for $n(T_F)$ from values of the literature [19, 23] on silica glasses having the closest composition. Values are reported in table 1.

A law for refractive index temperature dependence for silica glass has been suggested by Polian [10], from the comparison between Brillouin shift and ultrasonic measurements [11] but high frequencies and low frequencies sound velocity can have different values if relaxation takes place. As in this work we want to study relaxations effects in the glass transition range, we use instead, a refractive index thermal dependence measured and presented by Bruckner [19] for a type II silica glass : n(T) grows from 273 to 1273K-almost linearly following equation 7.

$n(T)=n(room\ temperature) +T*18.6*10^{-6}+0.6*10^{-8}*T^2-0.0002$ (7)

We have extrapolated this law for all the measured temperatures. No experimental measurement of the value of n(T) above 1273K reported in literature can ascertain the extrapolation, but the continuous evolution of n(T) up to 2000K is in adequacy with the reference [10].

Haken reports slight modifications of refractive index induced by OH, Cl, and F impurities [33]. Actually, as OH groups in silica glass are expected to break some Si-O-Si bridges, the number of non bridging oxygen (having a higher polarizability) is expected to increase when OH concentration increases, thus refractive index is expected to increase slightly with increasing OH content. In the case of samples of different nature, found in the literature, refractive index is not always known with accuracy, then an error on the determination of the refractive index has to be considered. We used an error on the refractive index $\Delta n=5*10^{-4}$ which leads to an additional error estimated around 2m/s for the sound velocity.

*2.3. Samples*

The first set of samples (table 1.c) is composed by low OH content (some ppb) silica glasses A (type I) which have been heat treated at temperatures $T_a$ ranging from 1373 to 1773K, and quenched fast enough in order to freeze-in in the new structural organization. The samples treatments have been detailed in a previous paper [46]. Apart from the heat treatment achieved at 1373K during six days, all the samples have been heat treated during 1h and 45mn. This treatment time, $t_a$, is longer than the stabilization time at those different temperatures. The samples structural evolution versus fictive temperature has been characterized various experimental techniques [46, 47, 51].

A second set of samples (table 1.a) containing silica glasses prepared by different methods was investigated. Two commercial samples have been widely studied : sample B, so called «Puropsil A», is a fused quartz provided by Quartz et Silice, produced from natural quartz by electrical fusion. Sample C, so called «Suprasil W» from Heraeus is synthetic silica produced from $SiCl_4$ in a water-free plasma flame. Two extra samples presented in [52] have also been measured at room temperature: sample E so called "GE-124" from General Electrics and Sample F so called "Corning 7980". Finally sample D is sol-gel silica, prepared using an aerogel of initial density $0.3 g.cm^3$, heat-treated at about 1373K for several hours until its density reaches the value $2.20$ $g.cm^3$ of amorphous silica. This sample corresponds to a higher OH content as well as a lower heat treatment temperature. The samples OH and impurities contents are reported in table 1a.

Some results will be compared to previous results from very low frequencies ultrasonic measurements by Fraser [50] on two samples which are also presented in table 1b.

**Table 1a.** Characteristics of the different commercial samples studied, their impurities content is given when known.

| Sample | A | B | C | D | E | F |
|---|---|---|---|---|---|---|
| Origin | GE | Puropsil A Quartz et silice | Suprasil W Heraus | Sol-Gel | GE 124 fused quartz | Corning 7980 |
| Class | I | I | IV | | I | III |
| [OH] | ppb | 20ppm | 0.4ppm | 3000ppm | 2ppm | 900ppm |
| Other impurities | Al | metallic | 200ppm Cl | | | Cl |
| density | 2.204 | 2.203 | 2.204 | 2.20 | | |

**Table 1b.** Characteristics of the commercial samples studied by Fraser [50]. Our results will be compared to the results shown in his low frequency ultrasonic study.

| Sample | Fraser [50] | Fraser [50] |
|---|---|---|
| Origin | IR vitreosil | Corning 7940 |
| Class | I | III |
| OH | 2.4 ppm | 900ppm |
| Other impurities | - | Cl |
| Density | 2.2024 | 2.2013 |

**Table 1c.** Characteristics of the heat treated samples A : densities measured by pycnometer and refractive index estimated from Bruckner [19] and [23] for samples having compositions close to that of sample A.

| Sample A | AS | A1100 | A1200 | A1300 | A1350 | A1450 | A1500 |
|---|---|---|---|---|---|---|---|
| Annealing temperature | As received | 1373K | 1473K | 1573K | 1623K | 1723K | 1773K |
| Density ( ±0.001) | 2,204 | 2.201 | 2.202 | 2.204 | 2.205 | 2.206 | 2.206 |
| Refractive index (300K, 514nm) | 1.4627 | 1.4624 | 1.4625 | 1.4626 | 1.4627 | 1.4628 | 1.4628 |

3. **Results**

## 3.1. Room temperature measurements: influence of fictive temperature

First, samples from set A having different fictive temperatures have been measured at room temperature in order to determine their longitudinal sound velocity. Fig 1 shows that longitudinal Brillouin peak increases when fictive temperature increases from 1373K to 1773K.

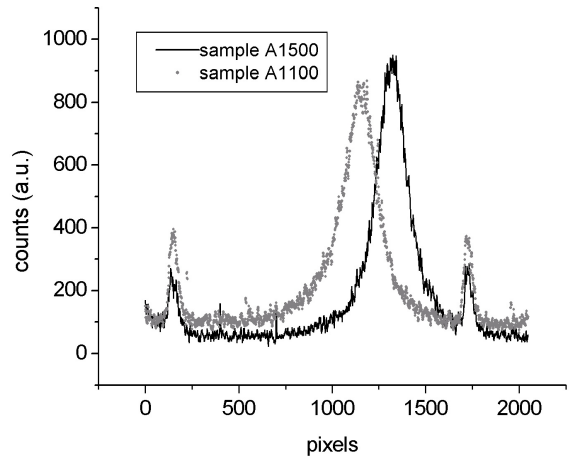

Figure 1 : Spectra of Brillouin scattering at room temperature for samples A silica glasses respectively heat treated at 1373 and 1773K.

Figure 2 shows longitudinal sound velocity versus fictive temperature for samples A annealed from 1173K to 1773K. The evolution of the refractive index n with fictive temperature ( table 1a.) has been taken into account for the calculation of $V_L$. It appears that for silica glass $V_L$ increases linearly by $8.8*10^{-2}$ $m.s^{-1}.K^{-1}$ with increasing fictive temperature in the range 1173K to 1773K -apart from the sample having a fictive temperature equal to 1773K, probably associated to an artefact or constraints-. Using this linear relationship, the fictive temperature of the as received sample A, is thus found equal to 1653±10 K. This result is consistent with the previous Raman and IR estimation equal to 1643±40K [46, 47].

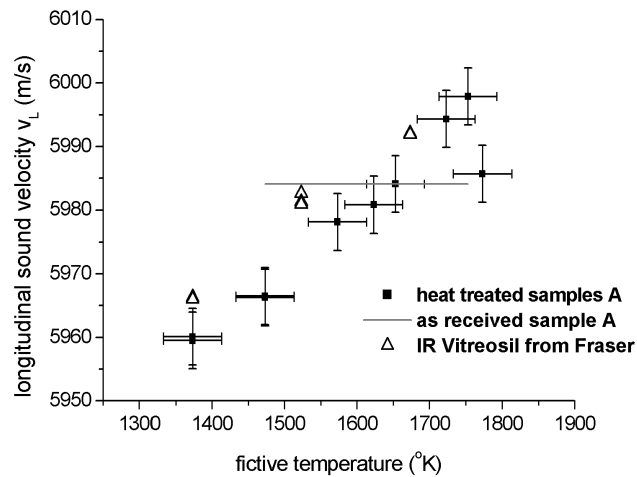

Figure 2 : Room temperature longitudinal sound velocity for sample A (■) versus fictive temperature measured by high resolution Brillouin spectroscopy . Full line gives the position of as received A sample. The ultrasonic results from Fraser [50] on a silica glass containing a low OH concentrations (table 2) are also presented for comparison on this figure and will be referred to in discussion.

Low frequency ultrasonic measurements results from Fraser [50] on silica samples containing few OH groups (IR Vitreosil) -similar to sample A- are also shown on figure 2. It must be noted that difference in frequency between Fraser and our measurements can partly account for the sound velocity difference [53, 54]. For IR Vitreosil $V_L$ and for sample A, fictive temperature dependences are parallel.

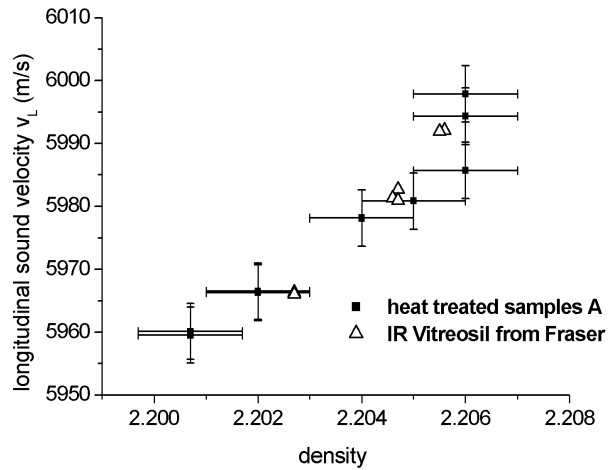

Figure 3 : Room temperature longitudinal sound velocity versus sample density for sample A (■) and for IR Vitreosil measured by Fraser [50] (table 1). Differences in density occur from variation in fictive temperature rather than from variations in OH concentration (IR Vitreosil has 2.4ppm OH and sample A contains few ppb OH).

Fig 3 shows longitudinal sound velocity plotted this time as a function of density for different heat treated samples: samples A from this study and IR Vitreosil from Fraser [50]. It appears that almost independently of the samples nature, $V_L$ evolves with density with a variation close to $7.2*10^{-3}\,m^4.s^{-1}g^{-1}$. Then, $V_L$ increases almost linearly with both fictive temperature and density which is consistent with the idea of a linear increase of density with $T_F$ (below 1673K).

*3.2. In situ temperature measurements: influence of fictive temperature*

Fig 4 shows longitudinal sound velocity variations with temperature across the glass transition interval. Two different analyses of Brillouin shift data as a function of temperature are compared. In most data of the literature n is taken independent of T. However, it is important to consider the evolution of n with temperature [19] because differences between the two analysis are significant in particular at high temperature (at 1773K the difference between the two analysis is of the order of 175m.s$^{-1}$ whereas the sound velocity variation in the range 300-1773K is around 540m.s$^{-1}$).

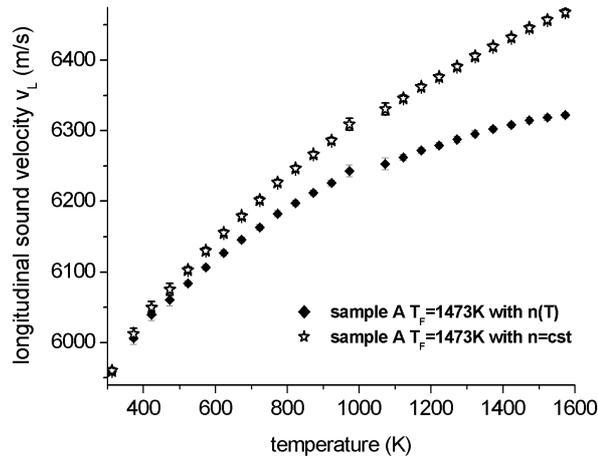

fig 4 : Sound velocity versus temperature for sample A : comparison between an analysis with n(T) and an analysis n=constant.

Figure 5 shows comparison of the temperature dependence of the longitudinal sound velocity in the range 300-1773K for two samples A, having respectively fictive temperatures equal to 1773K and 1473K. Measurements have been performed while heating with a rate about 1.30±0.05K/mn whereas spectra are measured during isothermal annealing times within 10mn. We considered for this calculation that n(T) increases with temperature, following the thermal dependence interpolation from [19]. A continuous increase is observed for $V_L$ but the slope becomes smaller at high temperature. The sample having a higher fictive temperature has a higher sound velocity up to approximately 973K, then the two curves join together and a small difference in sound velocity is again observed from 1273 to 1573K. The difference between the two curves has been plotted on fig 6 to emphasize the influence of T and $T_F$. The bump in the range 1273-1773K may be attributed to structural relaxation in the glass transformation range. However the difference at room temperature which decrease with increasing temperature and vanish at 973K is a new feature.

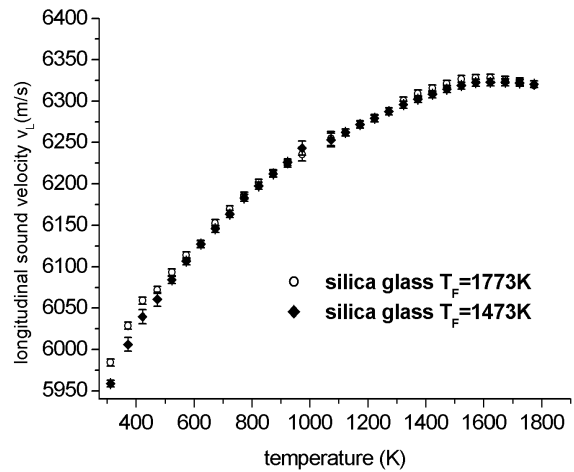

(a)

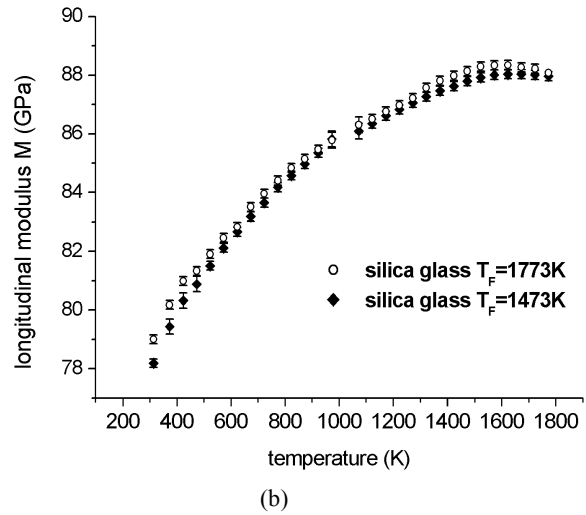

Figure 5 a and b : Longitudinal sound velocity (a) and longitudinal modulus (b) measured for two samples A heat treated at temperatures $T_a$ respectively equal to 1473K (diamond) and 1773K (open circle). Error bars are calculated taking into account error on shift determination and on scattering angle variation. Some larger error bars are related to a poorer Brillouin shift extraction due to a difficult deconvolution between the elastic line and the Brillouin line.

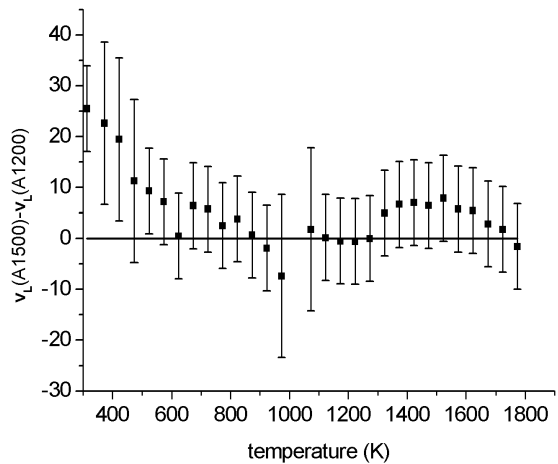

Figure 6: Longitudinal sound velocity difference between sample A, $T_F$=1773K and sample A $T_F$=1473K. The horizontal line is a guide for the eyes giving a difference equal to zero.

In order to show that structural relaxation effect gives rise to longitudinal sound velocity variations, we have measured in-situ isothermal evolution of longitudinal Brillouin shift at the annealing temperature 1448K (figure 7) as a function of time, after fast cooling of the sample from 1773K. The stabilization time found equal to 400min at 1448K, lies in the interval [105-4320] min corresponding respectively to the heat treatment times used for the stabilization of samples A1200 and A1100. Although the changes are small (about 6m/s) almost of the order of the experimental error the data can be well fitted by a decreasing exponential law, in agreement with a relaxation process with a characteristic time equal to 156mn.

The same in-situ relaxation measurement has also been performed at 1448K by SAXS [42, 55]. The density fluctuations relaxation has also been fitted using an exponential decay and give a characteristic decay time equal to 36mn. So the characteristic decay time for sound velocity seems much longer than the decay time for density fluctuations during isothermal relaxation at the same temperature. This difference can be explained, apart from a possible difference in the measurements of the furnace temperature (the highest error on temperature can be found equal to 5K), by the different length scales associated with the two experiments (Brillouin scattering is a macroscopic probe whereas SAXS probe length scales corresponding to the medium range order).

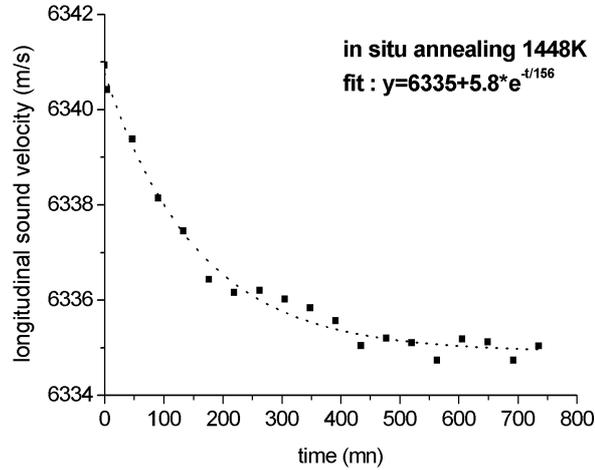

Figure 7: longitudinal sound velocity measured isothermally at 1448K versus time for a sample fast cooled from 1773K to erase its thermal history.

*3.2. Influence of the OH content*

The second part of the results concerns the influence of hydration on longitudinal and transverse sound velocity in silica glasses. Fig 8 shows room temperature longitudinal sound velocity $V_L$ versus OH content for different samples (B, C, D, E, F). Longitudinal sound velocity, $V_L$, tend to decrease with increasing OH content (figure 8), yet some differences for very low OH content may be due either to other impurities or to fictive temperature. For high OH content the amount of other impurities seem to have a negligible effect.

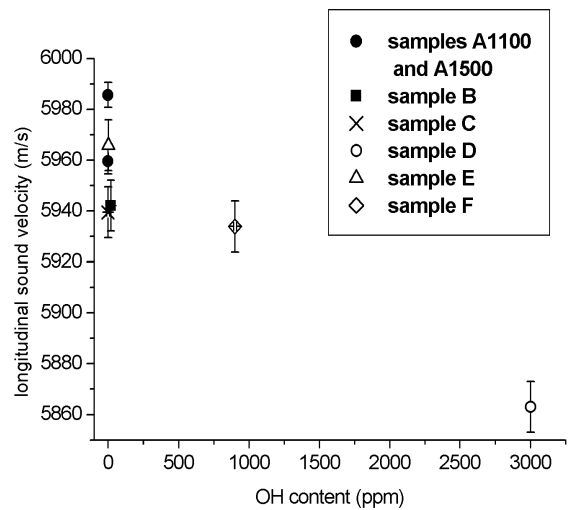

Figure 8: Room temperature longitudinal sound velocity versus OH concentration for some silica glasses presented in table 2. For samples A (full squares) two fictive temperatures have been shown in order to give a scale of variations due to OH as compared to variation due to fictive temperatures. Errors bars are taking into account a large uncertainty on n for samples B, C, D, E, F.

OH content is often related to relaxation processes when water diffusion is thermally activated [56]. Temperature dependence of the longitudinal and transversal sound velocity have been measured for the more hydrated (sol-gel glass) and two dehydrated silica glasses. Results for longitudinal sound velocity are shown on figure 9. Within the accuracy of the experiments, the longitudinal sound velocity temperature dependences are similar for the three samples apart from a slight kink which is observed in sol gel glass around 1173K. In the same figure, data from room temperature to 1773K are also compared to data from Polian [10] reanalyzed with the same refractive index temperature dependence taken from Brukner [19] than the one used for others samples. The thermal dependence found is slightly higher than the one observed for our samples. The comparison with the data from Polian [10] extends the temperature range of our measurements toward very high temperatures and let appear a small decrease of the sound velocity above 1800K, and the existence of a maximum value for $V_L$ between 1500K and 1800K.

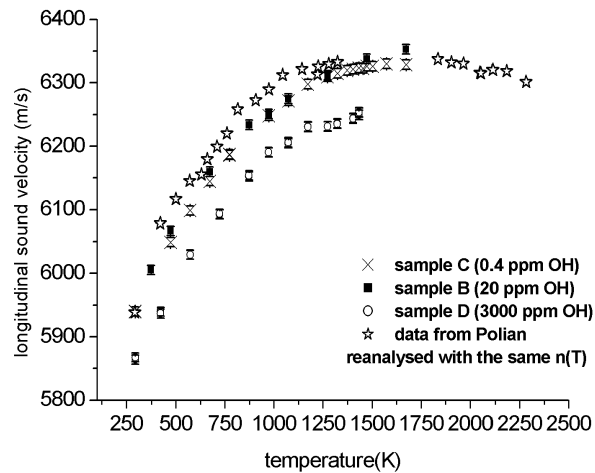

Figure 9 : Longitudinal sound velocity calculated from high resolution Brillouin scattering up to 1773K for three samples characterized by different OH concentrations (table 1.b). We also show a comparison with data from Polian [10] up to 2273K, reanalyzed with n(T) extracted from Brukner [19].

Transverse sound velocity for samples B and D are compared on figure 10. For the sol-gel silica, transverse sound velocity is about 35m.s$^{-1}$ whereas longitudinal sound velocity is 90m.s$^{-1}$ lower than for commercial sample B. A maximum is observed for the two samples, at 1373K for sample B and at 1173K for sample D. For temperatures above this maximum, transverse sound velocity starts to decrease for the two glasses, very clearly for the sol gel glass. The comparison with the data from Polian [10] reanalysed with n(T) used for the samples studied, confirms this decrease, provided n(T) can be extrapolated from [19] at temperatures as high as 2000K.

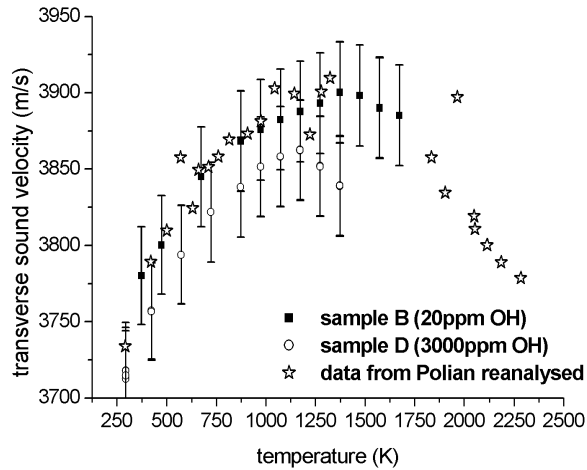

Figure 10 : Transverse sound velocity deduced from high resolution Brillouin scattering data up to 1773K for three samples characterized by different OH concentrations (table 1.b). We also show a comparison with data from Polian [10] up to 2273K, reanalyzed with n(T) extracted from Bruckner [19].

Longitudinal M and shear G moduli, can be deduced respectively from longitudinal and transverse sound velocity using the equations 3 and 4. Compression modulus K can then be calculated using equation 5. Figure 11 shows all three moduli calculated for samples B (Puropsil A) and D (sol gel). As for longitudinal and transverse sound velocity, bulk modulus temperature dependence for sample D and for sample B are parallel, the amplitude of the modulus for sample D is shifted by 3-5 % downward. The bulk modulus is always increasing with T even in the glass transition interval, whereas shear modulus decreases for temperatures higher than 1173K.

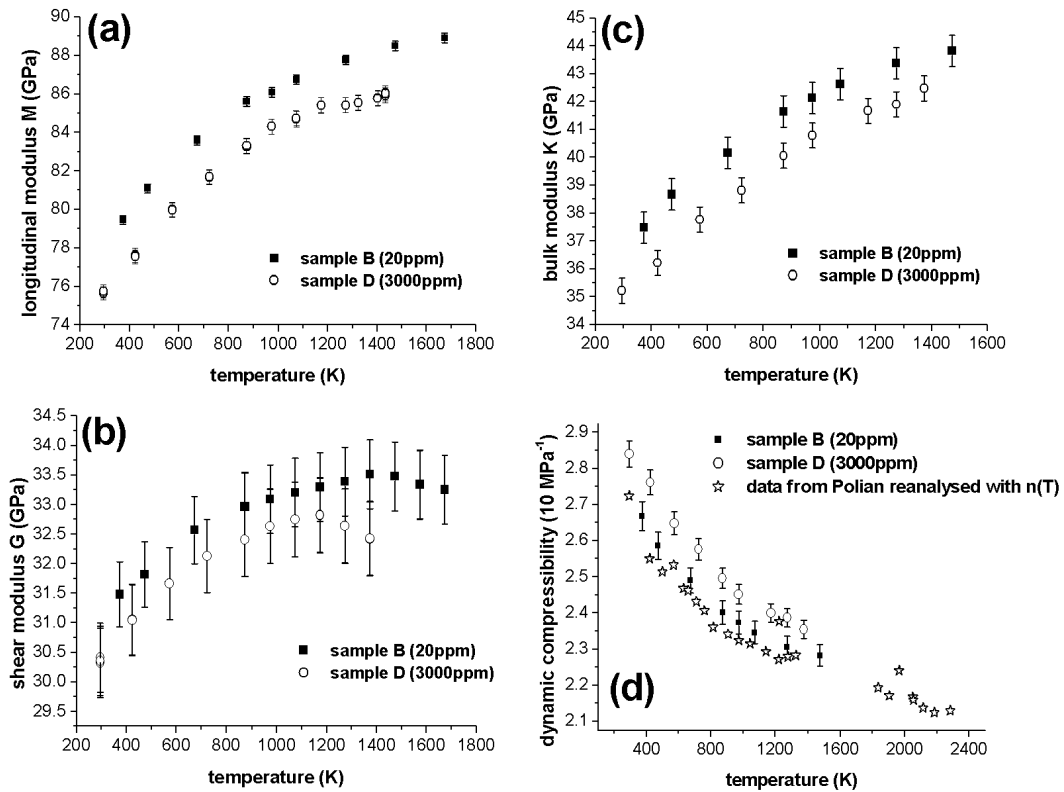

Figure 11 : longitudinal (a), transverse (b), bulk (c) moduli and adiabatic compressibility (d) for sample B (full squares) and D (open cicles) versus temperature.

Adiabatic compressibility at infinite frequency $\beta_\infty$ can be calculated from inverse compression modulus (equation 5) for Puropsil A (sample B) and for sol-gel silica (sample D), considering that the density of silica is almost constant with temperature. The variations of the modulus induced by the increase of the density with temperature are within the error bars and can be neglected. Fig 11 d shows results for the adiabatic compressibility for samples B and D versus increasing temperature. The comparison of the compressibility thermal dependence for the two commercial glasses with data from Polian [10] are in general agreement within experimental uncertainty, provided they are reanalyzed using the same refractive index temperature dependence [19].

**4. Interpretation**

*4.1. Sound velocity and moduli amplitude at room temperature : influence of fictive temperature and OH content*

In silica glass, we find that longitudinal sound velocity increases with increasing fictive temperature as a result of both Brillouin shift ($\Delta\nu=0.155$GHz- $\Delta\nu/\nu=0.45\%$ for $\Delta T_F=400$K) and refractive index ($\Delta n/n=0.03\%$ for $\Delta T_F=400$K) increase. This evolution of $V_L$ with $T_F$ for silica glass can be considered as anomalous if compared to alkali silicates, borate and phosphates glasses for which sound velocities decrease with increasing temperature. Few measurements, mostly Brillouin scattering measurements, are reported in the literature about the effect of thermal history on acoustic properties [34, 57, 58, 59, 60, 61]. For the different glasses reported in literature, sound velocity strongly decreases with increasing stabilization temperature.

Table 2 : Quantitative influence of fictive temperature on density $\rho$, refractive index n, longitudinal sound velocity $V_L$ and density fluctuations amplitude I(0) for different glasses.

|  | $\Delta T_F$ (K) | $(\Delta\rho/\rho)/(\Delta T_F/T_F)$ | $(\Delta n/n)/(\Delta T_F/T_F)$ | $(\Delta V_L/V_L)/(\Delta T_F/T_F)$ | $(\Delta I(0)/I(0))/(\Delta T_F/T_F)$ |
|---|---|---|---|---|---|
| $B_2O_3$ glass, dry [57] | [530-560]=30 | -3.0*10$^{-1}$ | - | -1.6 | |
| Float glass [55] | [753-893]=140 | -5.4*10$^{-2}$ | - | -1,0*10$^{-1}$ | 5.4*10$^{-1}$ |
| Sovirel glass [34] | [858-1023]=165 | - | -2.1*10$^{-2}$ | -2.3*10$^{-1}$ | |
| Silica sample A | [1373-1773]=400 | +8.9*10$^{-3}$ | +1.1*10$^{-3}$ | +2.6*10$^{-2}$ | 3.8*10$^{-1}$ |

Table 2 presents quantitative comparison of $(\Delta v_L/v_L)/(\Delta T_F/T_F)$ for different glasses as compared to relative ratio $(\Delta\rho/\rho)/(\Delta T_F/T_F)$ and $(\Delta n/n)/(\Delta T_F/T_F)$. It appears at first that variations in silica glass have an opposite sign as compared to variations in the three other glasses. Secondly, those three ratios are lower for glasses having increasing $\Delta T_F$. Then silica glass shows the smallest relative variation of longitudinal sound velocity but also the smallest relative density and refractive index variation versus fictive temperature. The variation of $V_L$ with $T_F$, in silica glass, are one order magnitude lower than in the other glasses and can be related with the very small slope for the thermal dependence of $V_L$ in the range 1373-1773K whereas for $B_2O_3$ and the other glasses the variation for $V_L$ in the glassy transition range is quite important [57].

We have shown that for silica glasses having different fictive temperatures, a linear relationship can be found between density and sound velocity. Such an evolution is not particular to silica glass: Ramos show that both $V_L$ and $V_T$ in borate glass having different thermal histories increase almost linearly with increasing density [57]. As a matter of fact, an increase of longitudinal sound velocity for different silica samples progressively densified under high pressure or permanently densified has also been observed [2 ,62]. Thus, for silica glass both sound velocity and density increase with increasing $T_F$, whereas for other silicate and borate glasses both sound velocity and density decrease with increasing $T_F$. It seems that anomalous behaviour of silica sound velocity is mostly related to the anomalous densification versus $T_F$.

The anomalous density behaviour in silica glass with increasing $T_F$ can be explained by the modification of the local and intermediate range structure. Local structural modification has been studied through vibrational IR and Raman spectroscopy. The shifts of the peak related to stretching vibration of the Si-O [31, 28-30, 46, 63] bonds, within Si-O-Si angles, are interpreted by a slight decrease of the intertetrahedral Si-O-Si angles θ distributed around a mean value. This decrease has been estimated (Δθ of the order of 2° for $\Delta T_F$ of the order of 400K) from Raman scattering [63] and from IR spectroscopy [27]. At intermediate range, silica glass structure can be described by the occurrence of rings containing from 3 to 9 tetrahedra. The two sharp peaks so called "defect lines" $D_1$ et $D_2$ observed in Raman spectrum at 495 $cm^{-1}$ and 606 $cm^{-1}$ have been associated with the vibrations of three and 4 membered rings. The intensification of the two lines with increasing fictive temperature-or with densifying under pressure or irradiation [63]-has been interpreted by a modification of those rings population [64], but this interpretation has been counterbalanced [65]. Molecular dynamics simulations for glasses obtained with different quenching rates also show a modification of ring statistics with quenching rate although fictive temperatures of simulated glass are very high as compared to experimental ones [65, 42]. This modification of ring statistics in favour of smaller ones would be consistent with a decrease of the mean Si-O-Si angle and thus a densification with increasing $T_F$.

Small composition effects such as OH content also influence silica glass properties such as acoustic properties (longitudinal and transverse sound velocities). Our results (figure 9) are consistent with ultrasound measurements: Kushibiki has shown that a sample containing 860ppm OH (and 30 ppm Cl) has its longitudinal and transverse sound velocities respectively 57m/s and 5m/s lower than for an OH free sample [66]. Both density and $V_L$ decrease with increasing OH content in silica glass through a relationship apparently linear. The decrease of the density with the incorporation of OH groups can be explained by the disruption of the continuous covalently bonded $SiO_4$ network

Finally, if fictive temperature effect and OH content effect are compared at room temperature, within the fictive temperature range and the OH content range available, the influence of fictive temperature ($\Delta v_L$=40m.s$^{-1}$ in the range of temperature experimentally available $\Delta T_F$=400°C) is less important than the influence of a high amount of hydroxyls ($\Delta v_L$=90m.s$^{-1}$ for $\Delta$[OH]=3000ppm). When OH content is low, the influence of other impurities such as Cl, is apparently not negligible anymore as shown by the amplitude of the sound velocity for sample C (Suprasil) being out of the general tendency. This sample is characterized by a lower elasticity than samples having the same OH content but Cl free. Impurities tend to modify the structure, in the case of OH and Cl a softening is observed.

*4.2. Sound velocity and moduli amplitude thermal dependence*

As for $T_F$ dependence, the temperature dependence of sound velocity in silica glass increases with increasing temperature. This positive slope, also reported in some other tetrahedral glasses such as $GeO_2$ (observed for all silica glass samples and all type of moduli below the glass transition range) is anomalous as compared to other usual glasses (borosilicate or float glasses).

$SiO_2$ and $GeO_2$ are both glass formers with highly connected networks. According to Kul'bitskaya [20], the difference between those glass formers and their alkali derived glasses can be explained by the increase the number of non bridging oxygen in multicomponent glasses, modifying oxides disturb the continuity of the network in favour of inhomogeneities regions. For example, in $(NaO_2)_x(SiO_2)_{1-x}$ the anomalous positive thermal dependence disappear for sodium oxide content x higher than 0.16 [60].

$B_2O_3$ is also a simple glass formers but does exhibit normal elastic properties [20, 57]. In fact, its structure consists of highly connected planar triangle network instead of highly tetrahedrally networked structure. Then it appears that the anomalous elastic moduli thermal dependence in silica and germania glasses are strongly correlated to the tetrahedral highly connected network and to the nature of the link between atoms [20].

The comparison of acoustic properties thermal dependence for two different fictive temperatures gives another insight on silica glass peculiar behaviour. Acoustic properties in usual glasses such as float or barium glass have the following behavior:

- Low fictive temperature glass and high fictive temperature glass have different sound velocities at room temperature [34, 55].
- When temperature increases for T<[Tg], $V_L$ thermal dependence appears as translated toward lower values when fictive temperature increases but with the same thermal dependence.
- For T>[Tg] thermal history effect vanishes, structures and properties are in equilibrium and $V_L$ is the same for the samples having different $T_F$.
- In the glass transition interval, singular effects, such as hysteresis or kinks [3-5, 34, 55] around Tg, are expected for samples having different thermal histories, even if those different thermal histories do not lead to measurable differences at room temperature. These singularities may appear when the rate of configurationnal change becomes of the order of magnitude of the heating rate, and traduce the attempt of the glass to adjust its molecular configuration as the temperature is changed.

In silica glass, fictive temperature influence on the moduli does not, in contrast to other glasses, give rise to a simple curve translation in the glassy range. The influence of fictive temperature can be observed through small amplitude differences below 873K and also in the glass transition region. Samples A1200 and A1500 are different at room temperature but the curves join as early as 873K and finally a small difference can be detected around Tg from 1373K to 1673K.

New remarks also come out from the comparison of acoustic properties thermal dependence for glasses containing very low and very high concentration of hydroxyls. For sol gel silica glass, the amount of hydroxyl is very high and the sample differentiates with a smaller value of sound velocity and in the glass transition range, $V_L$ shows a plateau and $V_T$ a maximum.

Actually for transverse sound velocity the maximum is observed in both sol gel glass and Puropsil respectively around 1173K and around 1373K. The existence of a maximum in the range [1373, 1773K] is confirmed while plotting data from Polian [10] up to 2300K reanalysed with refractive index temperature dependence. This maximum occurs in the glass transition range. It is consistent with a negative fictive temperature dependence measured by Fraser [13] for $V_T$. This decrease observed in the glass transition range on $V_T$ and G may be connected to a structural relaxation assisted by the hydroxyls. A structural softening may also contribute to the decrease of $V_L$. As a matter of fact, $V_L$ is linked in the same time to K (compressibility) and to G (shear) (equation 5). Then the plateau observed for $V_L$ results from a compressibility which keeps on increasing with temperature whereas shear modulus G is decreasing.

**5. Discussion**

*5.1. Comparison between density fluctuations and Kul'bitskaya's model[20]*
Fluctuations originating from the change of the dielectric constant (dielectric susceptibility and polarizability, electronic densities) associated with local density fluctuations can account for the scattering of electromagnetic waves in glasses. Then it seems rightful to consider a model which tries to give a semi-quantitative explanation of the positive temperature dependence of elastic modulus in silica glass trough the existence of spatial fluctuations like the model established by Kul'bitskaya [20]. This model assumes that glass is composed of inhomogeneities resulting from the freezing-in of natural fluctuations around the glass transition. It is demonstrated that, in an elastic inhomogeneous medium, an effective elastic modulus $\varepsilon_{eff}$ can then be written as the difference of two terms: an average modulus $\varepsilon_0$ and a correction term depending on the amplitude of those fluctuations $<\Delta^2>$ (equation 8).
$$\varepsilon_{eff} = \varepsilon_0 - <\Delta^2> \quad (8)$$
The temperature dependence of the sound velocity in silica glass and in float glass (for comparison) can be discussed in the frame of the equation (8) considering a possible correlation between the density fluctuations $(\Delta\rho)^2/\rho$ measured by SAXS in the same samples [42, 51, 69] and the fluctuation term $<\Delta^2>$ evoked in the equation (8).

It has been demonstrated experimentally by SAXS [42, 51, 52, 69] that density fluctuations increase in silica glass much more with fictive temperature (38%) than the average density (0.89%) for the same fictive temperature range (table 2, table 3). Then, if the fluctuation term $<\Delta^2>$ is considered to evolve the same way than the density fluctuation term, $(\Delta\rho)^2/\rho$, this term is expected to rise with the freezing-in temperature for both glasses. The decrease imposed by the term $-<\Delta^2>$ should induce a decrease of the effective moduli $\varepsilon_{eff}$, if the average modulus $\varepsilon_0$ is supposed to evolve like the average density of the glass. This result is inconsistent with the experimental observation showing that the sound velocity V increases with fictive temperature in silica glass.

For float glass, it can also be demonstrated that density fluctuation increase faster (54%, if the concentration fluctuations dependence on $T_F$ is neglected) than the average density (5.4%) decreases with fictive temperature (table 2, table 3). Then it can be assumed that the effective modulus fictive temperature dependence is determined by the term $-<\Delta^2>$. Thus the decrease of the effective modulus is consistent is consistent with the experimental observation.

Table 3 : Comparison of SAXS intensity I(q=0) and sound velocity for silica glass and float glass for samples heat treated at temperatures corresponding to equivalent viscosities.

|  | A1200 | A1500 | Sample E | Sample F | Float glass [55] | Float glass [55] |
|---|---|---|---|---|---|---|
| $T_F$ or [OH] | 1200°C | 1500°C | 2ppm | 900ppm | 480°C | 620°C |
| I(0) e.u./molecule | 20.6 | 23.3 | 20 | 18.2 | 11.3 | 13 |
| $V_L$(300K) m/s | 5985 | 5998 | 5972 | 5940 | 6930 | 5925 |

The thermal dependence of the effective modulus can also be discussed assuming the same correlation. The density fluctuations term increases with temperature. If variation of the average modulus $\varepsilon_0$ may be related to the thermal expansion coefficient, being very small for silica as postulated by Kul'bitskaya, we can conclude that $d\varepsilon_{eff}/dt \sim -d<\Delta^2>/dT$. Thus the effective modulus would decrease with temperature in silica glass whereas the purpose of Kul'bitskaya's model was to demonstrate the positive temperature dependence of the effective modulus (ie sound velocity) observed experimentally. For float glass the experimental negative thermal dependence is well described by the model.

The evolution of the elastic modulus versus OH content can also be discussed in the same framework. SAXS measurements have shown that an increase by 900ppm of the OH content induce a decrease by about 10% of $(\Delta\rho)^2/\rho$ [52] whereas the average density $\rho$ is expected to increase by 0.02%. It seems difficult then to explain the decrease of longitudinal sound velocity with increasing OH for glasses containing a high amount of OH such as the sol-gel silica.

To summarize, for silica glass, when the elastic fluctuation term and the density fluctuation are supposed to be correlated, the effective moduli do not evolve with temperature/fictive temperature in the way predicted by Kul'bitskaya.

Thus if the fluctuation term $<\Delta^2>$ in the model from Kul'bistkaya is responsible for the anomalous positive thermal dependence of the elastic moduli in silica glass, the origin of this term must be different from density fluctuations. In the thermal dependence of $V_L$ shown in the results for silica glass, the two glasses having different thermal histories join at quite low temperature. This effect seems to demonstrate that a relaxation of $<\Delta^2>$ or of $\varepsilon_0$ may occur at temperature much lower than the temperatures of the glass transition interval. These very low temperature are associated to high viscosities and also huge relaxation times, so these low temperature relaxations can hardly be associated to relaxations from structural entities such as frozen-in density fluctuations. In the following paragraph different hypothesis can be developed.

The fluctuation term in Kul'bitskaya's model could be correlated to elastic fluctuations which do not involve any density fluctuation. Such elastic fluctuations might be bound with residual stress or with fluctuations of the elastic strength of the Si-O bond without any change of the bond length. In the case of residual elastic stress, an increasing temperature should decrease the stress and thus the decreasing term $-<\Delta^2>$ is expected to induce an increase of the effective modulus. The low temperature relaxation, mentioned in the presentation of figure 5, could be linked such an elastic relaxation process.
Recent results from numerical simulations are consistent with the hypothesis of a heterogeneous nature of glass on dynamical point of view [70, 71]. These dynamic simulations in Lennard-Jones systems have shown that dynamic heterogeneities do exist, not only in the supercooled liquid as already established [72-74], but also in the glassy state for all temperatures. Mobile and immobile particles have been studied: both kind of particles form cluster. Results from Rossi [75] show the existence of an inhomogeneous cohesion (elastic constant heterogeneities) on the nm scale in glass. The existence of some nanoscaled regions of low and high cohesion had been proposed formerly by Duval [76, 77] in order to explain the origin of the Boson peak.

The model from Kul'bitskaya explains the elastic properties of usual glasses through a mean field approach. Usual glasses (like float glass) are often very complex glasses containing many components for which a mean field approach is suitable. As a matter of fact, we can suspect that even the most different configurations are not so far from the average configuration than they are in a glass like silica (for example in silica glass the angular distribution is ranging from 120 to 180°). Thus simple glass like silica or germania, have very rigid networks for which this mean field approach may be less suitable, and topological or structural models are then necessary.

*5.2 Glass polyamorphism*

Several references are found in literature pointing out a strong correlation between silica glass elastic properties and those of its crystalline counterparts. A change between 773 and 873K in the temperature dependence of the elastic moduli in silica glass has been reported in the derivative of Landau Placzec Ratio [6] and in the adiabatic compressibility around 900K [10]. This change has been interpreted like a reminder of the α-β quartz transition [6] (linked with the presence of some quartz crystallites associated with the fused quartz elaboration process of the samples). The change in the elastic moduli has not been observed in any of the silica glass studied, but the change of slope in the adiabatic compressibility can also be observed in sample B and D.

The concept of polyamorphism in silica glass has been introduced to illustrate the fact that silica glass could be described like a mixture of a finite number of energetically distinct structural states [78]. Early models are based on the assumption of the existence of two energetically distinct structural states, in order to explain the very low temperature specific properties of silica glass through the tunnelling systems between two energy wells (corresponding to the two structural states often associated to states of silica crystalline polymorphs) [1].

Cristobalite has always been a favoured archetype for structural model of silica glass (Si-O-Si angles, average ring size…) for many simulations. Huang and Kieffer [78] have studied amorphous transitions under varieties of thermomecanical conditions with molecular dynamic simulations and reproduce various anomalies of elastic properties in pressure or temperature dependence. Their simulations reveal that thermomechanical anomalies of silica glass are due to a localized reversible structural "transition" close to those observed in the α-β cristobalite phase transformation. This"transition" would involve Si-O-Si bond rotation and thus would take place without any change in the local order (bond length and angles, ring size distribution) but would affect ring geometry that become more symmetric under expansion. As α-cristobalite phase is characterized by a lower modulus than the β-phase, when the proportion of the angles undergoing a rotation into the β configuration (determined by Pressure and Temperature increase) the elastic modulus increases. The idea of tetrahedron rotations has already been proposed by Buchenau [78] in order to explain low frequency vibrational modes**.** Anomalous density behaviour has also been related to a transition from a Low Density Amorphous state to a High Density Amorphous state [48, 49].

The increase of the modulus with increasing fictive temperature could thus be explained by a larger number of local entities which have been undergoing the transformation into the β configuration. The β-cristobalite phase being a high temperature disordered phase, it seems consistent to associate this phase to a high fictive temperature glass. However, β-cristobalite phase has a lower density than α-cristobalite phase, it is thus expected that a higher number of β-cristobalite configurations decrease the density, which is in disagreement with the increase of density observed in silica glass when $T_F$ increases,.

As for the sound velocity when OH content increases, it has been demonstrated that OH plays in favour of local relaxation in glass [56], thus if a stressed region contains some OH bonds the relaxation process may happen rather through the OH relaxation than through the rotation process, thus locally the modulus may not increase as the rotation process is not involved.

**Conclusion**

High resolution Brillouin scattering was used to monitor the variation induced by fictive temperature and OH content, in elastic properties of silica glass at room temperature and versus increasing temperature up to and across the glass transition. The thermal dependence of the elastic moduli is found to be positive in the glassy range whatever the $T_F$ and the OH content is. Yet in the glass transition range, the most hydrated samples show a slight decrease of the transverse elastic moduli beyond 1223K. Those hydrated samples also show lower moduli. The positive temperature dependence of the elastic moduli in silica glass, whereas usual glasses show negative thermal dependence, belongs to the numerous anomalies reported for silica glass. Furthermore, we demonstrate in this study the increase of the sound velocity with increasing fictive temperatures in silica glass which also emphasize the anomalous character of silica glass. The results have been discussed through the phenomenological model of the fluctuations proposed by Kulbitskaya [20] and the structural model inspired by the α-β cristobalite transition proposed by Huang [78]. The description of the results does not apply well to silica glass by simply considering that the fluctuation term equal to the amplitude of the density fluctuation, an independent elastic term should probably considered. The structural description by localized reversible structural "transition/rotation" seems to give interesting clues to understand the influence of fictive temperature and OH content. Future improvement in the understanding of the link between the structural modifications and glass elastic properties is expected thanks to the involvement of simulation in the description of glass behaviour.

**Acknowledgments:**
The work has been carried out through the help of S. Clement in the preparation of the heating device and optical measurements. We acknowledge L. Vaccarini for involvement in measurements. We also wish to thank B.Ruffle and R.Vialla for helping in the set up of the Brillouin device, and R. Vacher for the apparatus conception. Finally, we thank R. Brüning for lending some silica glass samples.